\documentclass{aa}

\usepackage{graphicx}
\usepackage{wasysym}
\usepackage{amssymb}
\usepackage{amsmath}

\usepackage{graphicx} % for plots/pictures
\usepackage[latin1]{inputenc} % Encodage du texte en entr<E9>e (accents, etc...)

\begin{document}

\title{A precise modeling of Phoebe's rotation}

\author{L.~Cottereau \inst{1}, E.~ Aleshkina \inst{2} \and J.~Souchay\inst{1}}

%\offprints{L. Cottereau \email{laure.cottereau@obspm.fr}}

\institute{\inst{1} Observatoire de Paris, Syst\`emes de R\'ef\'erence Temps Espace (SYRTE), CNRS/UMR8630, Paris, France\\ \quad (Laure.Cottereau, Jean.Souchay@obspm.fr) \\ \inst{2} Main (Pulkovo) Astronomical Observatory of the Russian Academy of Sciences, Saint-Petersburg, Russia (aek@gao.spb.ru)}
\date{}

\abstract
{}
{Although the rotation of some Saturn's satellites in spin-orbit has already been studied by several authors, this is not the case of the rotation of Phoebe, which has the particularity of being non resonant. The purpose of the paper is to determine for the first time and with precision its precession-nutation motion }
{We adopt an Hamiltonian formalism of the motion of rotation of rigid celestial body set up by Kinoshita (1977) based on Andoyer variables and canonical equations. First we calculate Phoebe's obliquity at J2000,0 from available astronomical data as well as the gravitational perturbation due to Saturn on Phoebe rotational motion. 
Then we carry out a numerical integration and we compare our results for the precession rate and the nutation coefficients with pure analytical model.}
{Our results for Phoebe obliquity ($23^{\circ}95$) and Phoebe precession rate (5580".65/cy) are very close to the respective values for the Earth. Moreover the amplitudes of the nutations (26" peak to peak for the nutaton in longitude and 8" for the nutation in obliquity) are of the same order as the respective amplitudes for the Earth. We give complete tables of nutation, obtained from a FFT analysis starting from the numerical signals. We show that a pure analytical model of the nutation is not accurate due to the fact that Phoebe orbital elements $e$, $M$ and $L_S$ are far from having a simple linear behaviour.}
{The precession and nutation of Phoebe have been calculated for the first time in this paper. We should keep on the study in the future by studying the additional gravitational effects of the Sun, of the large satellites as Titan, as well as Saturn dynamical ellipticity.}

\maketitle

\section{Introduction}

Understanding the rotation of gravitationally interacting bodies with an ever increasing accuracy is an important challenge of celestial mechanics. Many authors have modelled with very good accuracy the rotation of the Earth as Woolard (1953), Kinoshita (1977). The other planets of the Solar system followed logically (see for instance Yoder, 1997; Rambaux, 2007). Today thanks to new and very precise observational astrometric techniques the dynamical study of celestial bodies extended to comets, asteroids and satellites (Meyer, 2008; Sinclair, 1977; Wisdom, 1984, 1987). 

Nevertheless a few studies have been made on the rotational motion of Phoebe, the ninth satellite of Saturn, since its discovery in 1899 by W.H Pickering. In 1905, F.Ross etablished that this motion is retrograde and gave for the first time the orbital elements refering to the mean equinox and ecliptic of date. Jacobson (1998) determined new orbital elements thanks to Earth based astrometric observations from 1904 to 1996. More recently Emelyanov(2007) elaborated ephemerides of Phoebe from 1904 to 2027.
Others studies have been made on the composition and the inertial parameters of Phoebe (Aleshkina et al., 2010). Phoebe is the only non-synchronous satellite of Saturn with rather well known physical parameters, as the moments of inertia. As Phoebe is a dissymetric body (large dynamical flattening and large triaxiality) it is interesting to study its rotation.

Here for the first time, we propose to determine the combined motion of precession and nutation of Phoebe considered as a rigid body. After presenting the model (Sect.\ref{2}), we describe the orbital motion of Phoebe by fitting the curves of the temporal variations of the orbital elements $a, e, M$ and $L_{s}$ with polynomial functions. Thanks to a fast Fourier transform (FFT), the large periodic variations around the mean elements are analysed and show that the orbital motion of Phoebe is far from being keplerian (Sect.\ref{3}). Then we determine the motion of precession and nutation of Phoebe by numerical integration using the value of the obliquity obtained in section \ref{4}. Finally, after quantifying the validity of the developments done by Kinoshita (1977) for the Earth, when applied to Phoebe, an analytical model is constructed. 

All our results are compared to the results obtained by Kinoshita for the Earth. We show in section \ref{6} that the numerical approach is significantly more accurate to determine the precession and the nutation motion of Phoebe than the analytical model on the opposite of the Earth for which a very good agreement is reached. This is due to the large periodic departure from the keplerian motion of the orbit of Phoebe. We note however that it gives a fairly good approximation for the precession which is sensitive to the average value.

\section{Theoretical equation of the precession-nutation motion}\label{2}

The present study of the rotation of Phoebe considered as a rigid body, is based on the theoritical framework set by Kinoshita (1977) who adopted an Hamiltonian formulation using Andoyer variables (1923) to evaluate the motion of the precession-nutation of the rigid Earth.
\begin{figure}[htbp]
\center
\resizebox{1.\hsize}{!}{\includegraphics{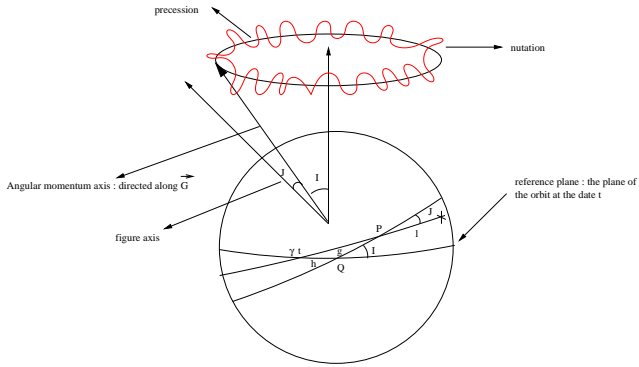}}
 \caption{The precession nutation motion}
\label{fig1}
\end{figure}
Here a simplified model is described and only the variables and the equations used in this paper are detailed. The rotation of a rigid Phoebe involves three axes : the figure axis, which coincides with the axis of the largest moment of inertia, the angular momentum axis directed along $\overrightarrow{G}$, with G the amplitude of the angular momentum and a reference axis arbitrarly chosen as the axis perpendicular to the orbit of Phoebe at a given epoch t (see Fig.\ref{fig1}). The precession and nutation motions are respectively the linear and the quasi-periodic parts of the motion of the figure axis or of the angular momentum axis with respect to the reference axis. Assuming that the angle $J$ between the angular momentum axis and the figure axis is small as it is the case of the Earth (for which its value is less than $1"$), only the motion of the angular momentum axis is considered here. The motion of this axis with respect to the reference axis is described by the angles $h, I$ (Andoyer,1923) where $I$ is the obliquity angle and $h$ is characterizing the precession-nutation in longitude which corresponds to the angle between the reference point $\gamma_{t}$ and the node $Q$. The reference point $\gamma_{t}$ is the intersection between the orbit and the equator of Phoebe at the date t , so-called "departure point" (Capitaine, 1986) and $Q$ is the ascending node between the plane normal to the angular momentum and the orbital plan. The Andoyer variables $g, l$ in the Fig.\ref{fig1} are respectively the angle between the node $Q$ and the node $P$ and the angle between a meridian origin and the node $P$ where $P$ is itself the ascending node between the plane normal to the angular momentum axis and the equatorial plane. The proper rotation of Phoebe is described by the angle $l+g=\Phi$. The Hamiltonian related to the rotational motion of Phoebe is:
\begin{equation}
K=F_{o}+ E+E'+U
\end{equation}
$F_{o}$ is the Hamiltonian for the free rotational motion defined by :
\begin{equation} \label{ham}
F_{0}= \frac{1}{2}(\frac{\sin ^{2} l}{A}+\frac{\cos ^{2} l}{B})(G^{2}-L^{2})+\frac{1}{2}\frac{L^{2}}{C}.
\end{equation}
where $L$ is the component of the angular momentum axis along the figure axis and $A, B, C$ are the principal moment of inertia of Phoebe. $E+E'$ is a component related to the motion of the orbit of Phoebe, which is caused by planetary perturbations and have been given in detail in Cottereau and Souchay (2009) when studying the rotation of Venus. $U$ is the disturbing potential due to Saturn considered as a point mass and its disturbing potential is given by :
\begin{equation*}
U = U_{1}+U_{2}
\end{equation*}
\begin{eqnarray}\label{eq1}
U_{1}&&=\frac{\mathtt{\textbf{G}} M'}{r^3}\Big[[\frac{2C-A-B}{2}]P_{2}(\sin \delta )\nonumber\\&&+[\frac{A-B}{4}]P_2^{2} (\sin \delta) \cos 2\alpha\Big]
\end{eqnarray}
\begin{eqnarray}
&U_{2}&=\sum_{n=3}^\infty \frac{\mathtt{\textbf{G}} M'M_{P}a^n}{r^{n+1}}[J_{n}P_{n}(\sin \delta)\nonumber\\&&-\sum_{m=1}^n{P_{n}^m (\sin\delta)\cdot (C_{nm}\cos m\alpha +S_{nm} \sin m\alpha) }]
\end{eqnarray}
where $\mathtt{\textbf{G}}$ is the gravitational constant, $M'$ is the mass of Saturn, $r$ is the distance between its barycenter and the barycenter of Phoebe. $\alpha$ and $\delta$ are respectively the longitude and latitude of Saturn (not to be confused with the usual equatorial coordinates), with respect to the mean equator of Phoebe and a meridian origin. The $P_{n}^m$ are the classical Legendre functions given by:
\begin{equation}\label{legendre}
P_{n}^m (x)= \frac{(-1)^m(1-x^2)^{\frac{m}{2}}}{2^n n!}\frac{d^{n+m} (x^2-1)^n}{d^{n+m}x}.
\end{equation}
Notice that the perturbations due to the other planets, satellites and the Sun a priori of second order, will not be studied in this paper. The variation of the obliquity and the precession angle is given by :
\begin{eqnarray}
&&\frac{dI}{dt}=\frac{1}{G}[\frac{1}{\sin I} \frac{\partial K}{\partial h}-\cot I\frac{\partial K}{\partial g}] \nonumber\\&&
\frac{dh}{dt}=-\frac{1}{G\sin I} \frac{\partial K}{\partial I} 
\end{eqnarray}
To solve these equations, Kinoshita (1977) used the Hori's method. As the order of the disturbing function is :
\begin{equation}
\frac{U_{1}}{F_{o}}\approx \frac{M'}{M'+M_{p}}(\frac{n}{\omega})^2(\frac{2C-A-B}{2C})\approx 3.46 \cdot 10^{-8}
\end{equation}
the same method can be applied to Phoebe. Neglecting the very small contribution as the component $E+E'$ and applying the first order of the Hori's method, this yields :
\begin{align}\label{eq2}
\Delta I =\frac{1}{G}[\frac{1}{\sin I} \frac{\partial}{\partial h} \int U_{1}dt-\cot I\frac{\partial}{\partial g}\int U_{1} dt]
\end{align}
\begin{align}\label{eq3}
\Delta h =-\frac{1}{G\sin I} \frac{\partial}{\partial I}{\int U_{1}dt}.
\end{align}
$ U_{1}$ given by (\ref{eq1}) can be expressed as a function of the longitude $\lambda$ and the latitude $\beta$ of Saturn with respect to the orbit of Phoebe at the date t using the transformations described by Kinoshita (1977) and based on the Jacobi polynomials:
\begin{eqnarray}\label{eq4}
U_{1}\nonumber&=&\frac{\mathtt{\textbf{G}} M'}{r^3}\Bigg[\frac{2C-A-B}{2}\Big(-\frac{1}{4}(3\cos ^2 I-1)\nonumber\\&&-\frac{3}{4}\sin^2 I \cos 2 (\lambda-h)\Big)\nonumber \\&& +\frac{A-B}{4}\bigg[ \frac{3}{2} \sin^2I \cos(2l+2g)\nonumber \\&& +\sum_{\epsilon=\pm 1}\frac{3}{4}(1+\epsilon \cos I)^2\cdot \cos 2(\lambda-h-\epsilon l-\epsilon g)\bigg]\Bigg].\nonumber \\
\end{eqnarray}
Finally the motions of precession-nutation in longitude and in obliquity are given starting from (\ref{eq2}), (\ref{eq3}) and (\ref{eq4}) by
\begin{eqnarray}\label{eq5}
\Delta h& =&-\frac{1}{\sin I} \frac{\partial}{\partial I} \int (W_{1}+W_{2}) dt
\end{eqnarray}
\begin{eqnarray}\label{eq6}
\Delta I &=&\Bigg[\frac{1}{\sin I} \frac{\partial}{\partial h} \int (W_{1}+W_{2}) dt \nonumber \\&& -\cot I \frac{\partial}{\partial g} \int W_{2} dt \Bigg]
\end{eqnarray}
where
\begin{eqnarray}\label{eq7}
&W_{1}&=\frac{\mathtt{\textbf{G}}M'}{Gr^3}\Big(\frac{2C-A-B}{2}\big[-\frac{1}{4}(3\cos ^2 I-1)\nonumber\\&&-\frac{3}{4}\sin^2 I \cos 2 (\lambda-h)\big]\Big)
\end{eqnarray}
\begin{eqnarray}\label{eq8}
W_{2}&=&\frac{\mathtt{\textbf{G}}M'}{Gr^3}\bigg(\frac{A-B}{4}\Big[\frac{3}{2} \sin^2I \cos(2l+2g)\nonumber \\ &&+\sum_{\epsilon=\pm 1}\frac{3}{4}(1+\epsilon \cos I)^2 \nonumber \\&& \cos 2(\lambda-h-\epsilon l-\epsilon g)\Big]\bigg).
\end{eqnarray}
Since $W_{1}$ does not depend on $g$, this variable does not appear in the second part of the right hand side of equation (\ref{eq6}). Supposing that the components $\omega_{1}$ and $\omega_{2}$ of the rotation of Phoebe are negligible with respect to the component $\omega_{3}$ along the figure axis, as it is the case for the Earth, this yields :
\begin{eqnarray}\label{eq9}
&W_{1}&=\frac{3\mathtt{\textbf{G}}M'}{\omega r^3}\Big(\frac{2C-A-B}{2C}\big[-\frac{1}{12}(3\cos ^2 I-1)\nonumber\\&&-\frac{1}{4}\sin^2 I \cos 2 (\lambda-h)\big]\Big)
\end{eqnarray}
\begin{eqnarray}\label{eq10}
W_{2}&=&\frac{3\mathtt{\textbf{G}}M'}{\omega r^3 }\bigg(\frac{A-B}{4C}\Big[\frac{1}{2} \sin^2I \cos(2l+2g)\nonumber \\ &&+\sum_{\epsilon=\pm 1}\frac{1}{4}(1+\epsilon \cos I)^2 \nonumber \\&& \cos 2(\lambda-h-\epsilon l-\epsilon g)\Big]\bigg).
\end{eqnarray}
where $\frac{2C-A-B}{2C}$ and $\frac{A-B}{4C}$ are respectively the dynamical flattening and the triaxiality of Phoebe. As the rotation of Phoebe is retrograde the conventional notations of the precession and obliquity will be respectively :
\begin{equation}\label{eq11}
\Delta \psi= \Delta h
\end{equation}
\begin{equation}\label{eq12}
\Delta \epsilon = \Delta I,
\end{equation}
which are the opposite of the conventional ones for the Earth (Kinoshita, 1977, Souchay et al., 1999).

\section{Phoebe Ephemerides}\label{3}

The ephemerides used in this paper for the Phoebe orbital motion around Saturn have been constructed by Emelyanov (2007) where the orbital parameters and the coordinates (X, Y, Z) of Phoebe are given in the planeto-equatorial reference frame considered as inertial from the first observation date in 1904 to 2027. The reference point is one of the intersections between Saturn and Earth equatorial planes at J2000.0. To apply the theoretical framework of Kinoshita (1977) the mean elements for the variables $a, e, M$ and $L_{s}$ are needed. In this section, these elements are calculated in the same way as Simon et al. (1994) for the planets. The differences betwen the true motion of Phoebe and a keplerian motion are also studied. Throughout the paper our study is restricted to the time span ranging from 2000.0 to the limit of the ephemeris 2027.0. 

\subsection{The semi-major axis}

\begin{figure}[htbp]
\center
\resizebox{0.6\hsize}{!}{\includegraphics[angle=-90]{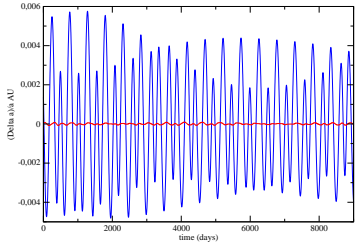}}
 \caption{Relative variation of $a$ around the mean value $\bar{a}=0.0864273$ in 9000 days time span.The central curve represents the residual after substracting the sinusoidal terms of the Table \ref{tab1new}.}
\label{fig3new}
\end{figure}

The mean element for the semi-major axis for the Earth is a constant (Simon et al, 1994) and the relative variations around this mean value are of the order $10^{-5}$. In comparaison, for Phoebe, approximating $a$ by a constant does not yield a good accuracy. Fig.\ref{fig3new} shows the relative variation of $a$ around its mean value estimated at $\bar{a}=0.0864273$ in 9000 days time span. The variations consist clearly in periodic components with amplitude of the order of $10^{-3}$. The leading oscillations of the signal are determined thanks to a fast Fourier Transform (FFT). Table \ref{tab1new} gives the leading amplitudes and periods of the sinusoids characterizing the signal in Fig.\ref{fig3new} where the central curve represents the residuals after substraction of these sinusoids.

\begin{table}[!htpb]
\caption{Leading amplitudes and periods characterizing the semi-major axis.}
\label{tab1new}
\begin{center}
\resizebox{0.5\hsize}{!}{\begin{tabular}[htbp]{rrr}

\hline \hline
 Period & amplitude&amplitude \\
  &sin&cos\\
   days &AU & AU\\
\hline\\
261.98&-2.319 $10^{-4}$&2.609 $10^{-4}$\\

498.93&-3.309 $10^{-5}$&-1.756 $10^{-4}$ \\

177.54&-5.715 $10^{-5}$&-9.971 $10^{-6}$\\

\hline
\end{tabular}
}
\end{center}
\end{table}
The presence of the relative large periodic components due to the additional perturbations on the orbit of Phoebe shows that the motion of this satellite is not close to a keplerian one, as it is the case for the planets of the Solar system. The effects of the attraction of the Sun, Jupiter and the other satellites of Saturn on Phoebe 's orbit entail an important departure from the keplerian motion. The individual study of the other orbital elements confirms this result.

\subsection{The mean elements of the orbital parameters}
\begin{figure}[htbp]
\center
\resizebox{0.6\hsize}{!}{\includegraphics[angle=-90]{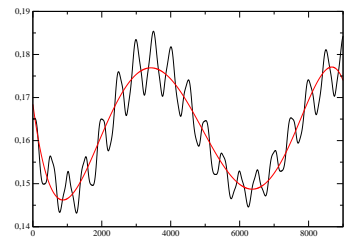}}
 \caption{Variation of the eccentricity of Phoebe for a 9000 days time span. The central curve represents the polynomial function given by (\ref{e}).}
\label{fig18}
\end{figure}

In the following, the formula for the mean elements of Phoebe keplerian motion $e, M$ and $L_{S}$ are given, where $M,L_{s}$ are respectively the mean anomaly and mean longitude with respect to the "departure point" $\gamma_{t}$. They were obtained by fitting the curves of the temporal variations of these elements given by Emelyanov (2007) with a polynomial expression at order 6 as it has been done for the planets of the Solar system (Simon et al, 1994). Considering the limits of the ephemerides, truncating the polynomial functions at the order 6 looks as a sufficient approximation. Thus : 
\begin{eqnarray}\label{e}
e&\approx &0.168480-0.0000594 \cdot t+ 4.931508\cdot 10^{-8}\cdot t^2\nonumber\\&&
-1.30162\cdot 10^{-11} \cdot t^3+ 1.1744\cdot 10^{-15} \cdot t^4 \nonumber\\&&
-5.2502\cdot 10^{-22} \cdot t^5-3.0861\cdot 10^{-24} \cdot t^6,
\end{eqnarray}
\begin{eqnarray}\label{mp}
M&\approx&58.9233+0.656386 \cdot t-0.000017\cdot t^2\nonumber\\&&
+9.78065\cdot 10^{-9} \cdot t^3-2.1541\cdot 10^{-12} \cdot t^4\nonumber\\&&
+2.06592 \cdot 10^{-16} \cdot t^5-7.2487 \cdot 10^{-21} \cdot t^6
\end{eqnarray}
\begin{eqnarray}\label{ls2}
L_{s}&\approx&331.109+0.653712 \cdot t-3.64487\cdot 10^{-7}\cdot t^2\nonumber\\&&
+3.54186\cdot 10^{-10} \cdot t^3-9.42668\cdot 10^{-14} \cdot t^4\nonumber\\&&
+1.03032 \cdot 10^{-17} \cdot t^5-4.02263 \cdot 10^{-22} \cdot t^6
\end{eqnarray}
where $t$ is counted in julian days, $M$ and $L_{s}$ being expressed in degrees. Fig.\ref{fig18} shows the variations of the eccentricity of Phoebe for a 9000 days time span. An important feature of Phoebe orbit is the large eccentricity as well as its very large relative variations (of the order of $10 \%$). As we will see in section \ref{6} this will be very important in our study because of the large corresponding variations of $(\frac{a}{r})^3$ that it causes for the orbit.

\begin{figure}[htbp]
\center
\resizebox{0.6\hsize}{!}{\includegraphics[angle=-90]{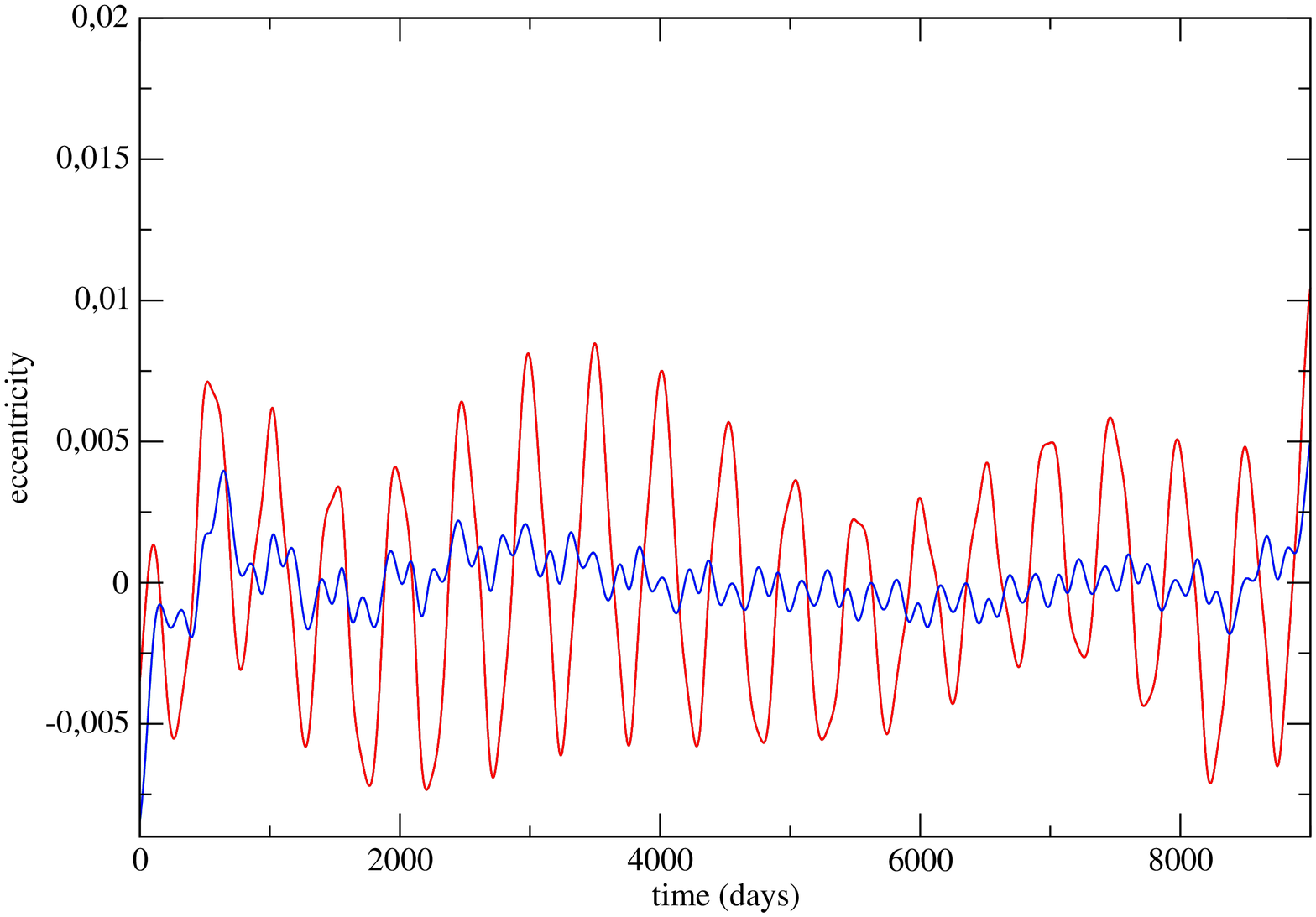}}
 \caption{Residuals after substraction of the polynomials given by (\ref{e}) from the eccentricity of Phoebe for a 9000 days time span. The central curve represents the residuals after substracting the sinusoidal terms of Table \ref{tab2new}.}
\label{fig20}
\end{figure}

\begin{figure}[htbp]
\center
\resizebox{0.6\hsize}{!}{\includegraphics[angle=-90]{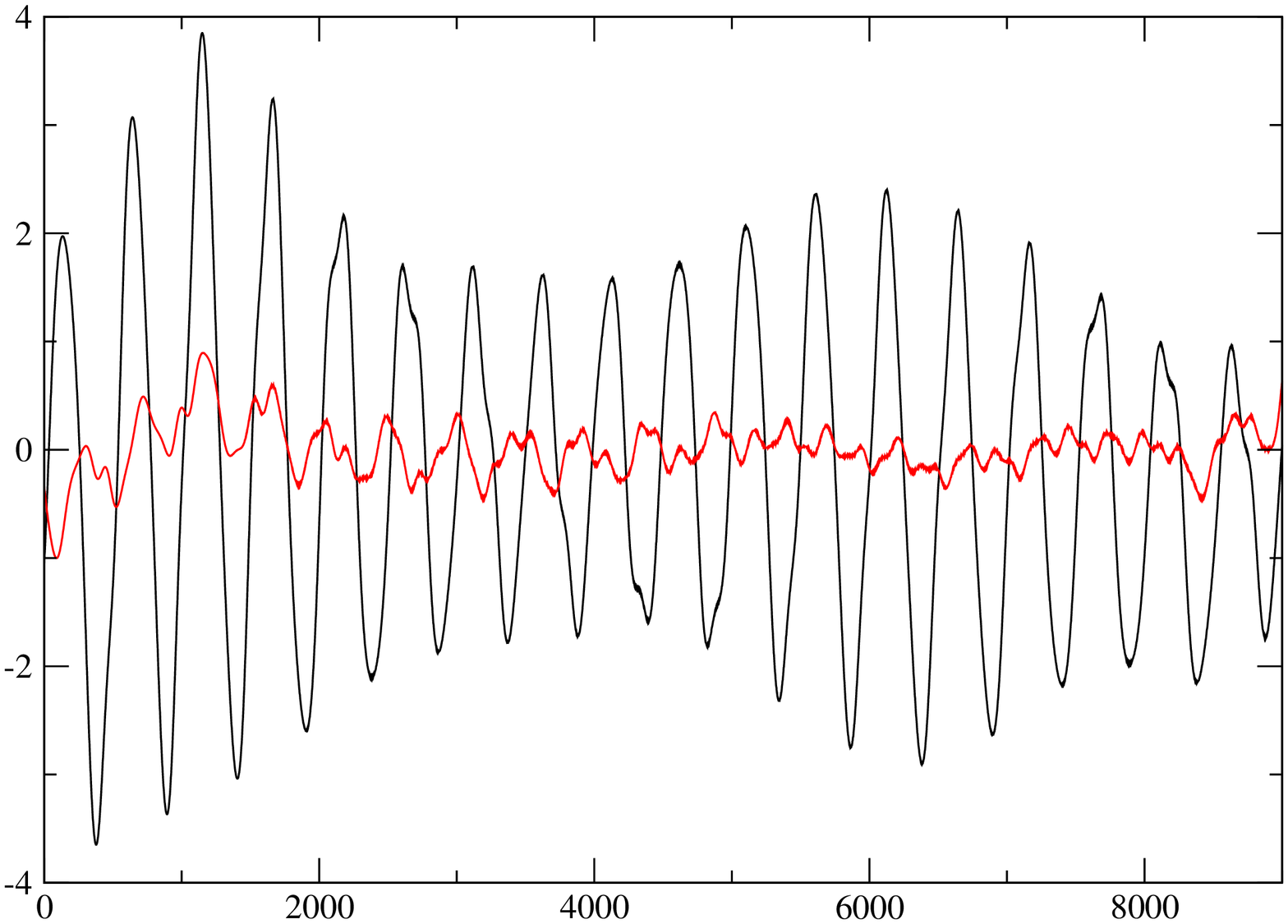}}
 \caption{Residuals after substraction of the polynomials given by (\ref{mp}) from the mean anomaly of Phoebe $M$ for a 9000 days time span. The central curve represents the residuals after substracting the sinusoidal terms of Table \ref{tab2new}.}
\label{fig14}
\end{figure}

\begin{figure}[htbp]
\center
\resizebox{0.6\hsize}{!}{\includegraphics[angle=-90]{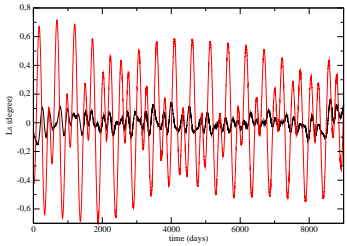}}
 \caption{Residuals after substraction of the polynomials given by (\ref{ls2}) from the mean longitude of Phoebe $L_{s}$ for a 9000 days time span. The central curve represents the residuals after substracting the sinusoidal terms of Table \ref{tab2new}.}
\label{fig17}
\end{figure}

Figures \ref{fig20}, \ref{fig14} and \ref{fig17} represent respectively the residuals after substraction of the polynomial functions given by (\ref{e}), (\ref{mp}) and (\ref{ls2})from the eccentricity, the mean anomaly and the mean longitude of Phoebe. As already observed for the semi-major axis, the residuals have periodic components which reach an amplitude of $4^{\circ}$ for the mean anomaly. 

\subsection{Frequencies analysis}

Thanks to a FFT analysis the leading ampitudes and the periods of the sinusoids characterizing the signals in Figs.\ref{fig20}, \ref{fig14} and \ref{fig17} are determined and given in Table \ref{tab2new}.

\begin{table}[!htpb]
\caption{Leading amplitudes and periods characterizing the orbital elements $e, M$ and $L_{s}$.}
\label{tab2new}
\begin{center}
\resizebox{0.5\hsize}{!}{\begin{tabular}[htbp]{crrr}

\hline \hline
 orbital element&Period & amplitude&amplitude \\
  &&sin&cos\\
   &days & & \\
\hline\\
&501.078&7.1200 $10^{-4}$&5.776 $10^{-3}$\\

eccentricity&3505.99&1.4949 $10^{-3}$&8.19196 $10^{-4}$ \\

e&550.62&1.260 $10^{-3}$&-8.9235 $10^{-4}$\\
\hline\\
&498.88&2.4850&-0.488\\

Mean anomaly&553.38&0.5212&0.2515\\

M&177.50&1.20$10^{-2}$&-0.2338\\
\hline\\
&262.02&-0.2959&-0.2669\\

Mean longitude&499.25&0.3470&-5.2116$10^{-2}$\\

$L_{S}$&554.2945&6.7272$10^{-2}$&7.1011$10^{-4}$\\
\hline
\end{tabular}
}
\end{center}
\end{table}

The central curves in Figs.\ref{fig20}, \ref{fig14} and \ref{fig17} represent the residuals after substraction of the leading sinusoidal components of the mean elements. We want to highlight here the fact that the motion of Phoebe is far from being a quasi keplerian motion as it is the case for the planets of the Solar system. In the following the precession-nutation of Phoebe is determined both analyticaly and numericaly. The differences between the results given by the two methods due to the non keplerian motion of Phoebe will be significantly increased when compared with the equivalent study for the Earth (Kinoshita, 1977). All the Phoebe parameters involved in the equations (\ref{eq5}) and (\ref{eq6}) which enables one to calculate the precession and nutation are known, at the exception of the obliquity of Phoebe at J2000.0, which represents a fundamental initial condition. It is calculated in the following section.

\section{Obliquity of Phoebe at J2000.0}\label{4}

Using a similar method as in Cottereau and Souchay (2009), the obliquity is given by :
\begin{equation}\label{eq13}
\cos I_{0} = \overrightarrow{P_{op}} \cdot \overrightarrow{P_{p}} 
\end{equation}
where $P_{op}$ and $P_{p}$ are respectively the unit vectors directed toward the orbital pole and the pole of rotation of Phoebe at $t_{0}$, defined by :
\begin{eqnarray}
\overrightarrow{P_{op}}=\begin{pmatrix}
\sin i_{0} \sin \Omega_{0}\\ -\sin i_{0} \cos \Omega_{0} \\ \cos i_{0}
\end{pmatrix}
\end{eqnarray}
\begin{eqnarray}
\overrightarrow{P_{p}}=\begin{pmatrix}
\cos \delta_{p}^{0} \cos \alpha_{p}^{0} \\ \\ \sin \alpha_{p}^{0} \cos \delta_{p}^{0} \\ \\ \sin \delta_{p}^{0}
\end{pmatrix}
\end{eqnarray}
$i_{0}$, $\Omega_{0}$ are the inclination and the longitude of the ascending node of Phoebe in the Saturn equatorial reference frame at J2000.0 given by the ephemerides of Emelyanov (2007) and $\alpha_{p}^{0}$, $\delta_{p}^{0}$ are the geo-equatorial coordinates of the north pole of Phoebe given by Seidelmann et al. (2007). The numerical values of these variables are given in appendix. Applying two rotations to $\overrightarrow{P_{p}}$, the two units vectors are expressed in the Saturn equatorial reference frame and the obliquity at J2000.0 can be determined directly from (\ref{eq13}).

After computation and taking into account the Phoebe retrograd rotation the numerical value of the obliquity at J2000.0 is :
\begin{equation}
I_{0}=23^{\circ}.95
\end{equation}
Notice that this obliquity is close to the Earth one ($23^{\circ}.43$). Knowing the obliquity at J2000.0, the precession-nutation of Phoebe can be determined with the help of (\ref{eq5}) and (\ref{eq6}) in the following sections.

\section{Numerical results of the precession-nutation of Phoebe}\label{5}

\begin{figure}[htbp]
\center
\resizebox{0.7\hsize}{!}{\includegraphics[angle=-90]{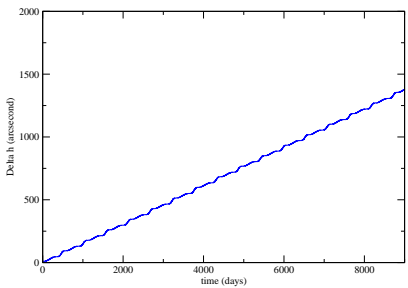}}
 \caption{The precession and the nutation of Phoebe in longitude for a 9000 days time span.}
\label{fig3}
\end{figure}

\begin{figure}[htbp]
\center
\resizebox{0.7\hsize}{!}{\includegraphics[angle=-90]{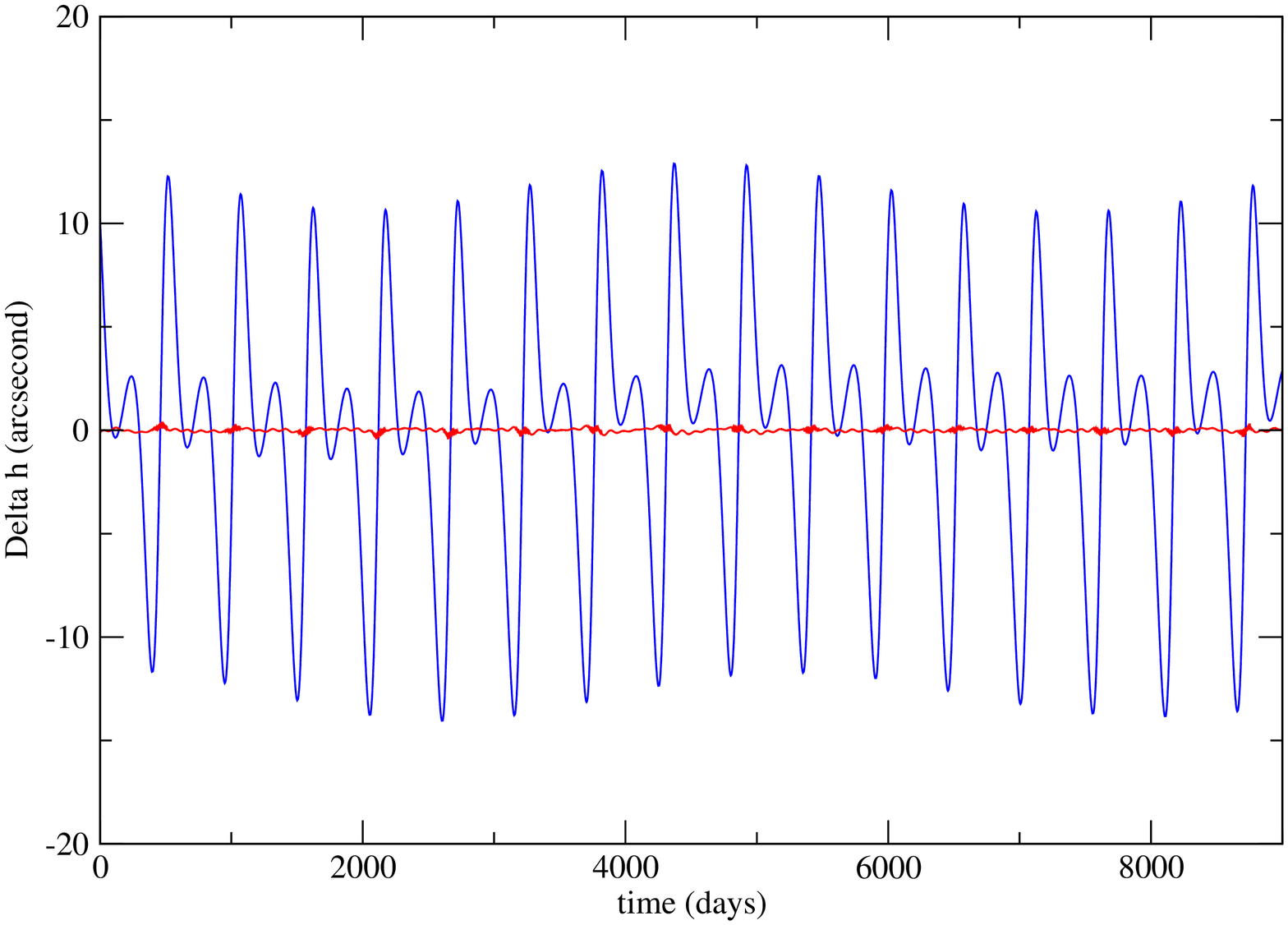}}
 \caption{The nutation of Phoebe in longitude for a 9000 days time span.The central curve represents the residual after substracting the sinusoidal terms of the Table \ref{tab1}.}
\label{fig4}
\end{figure}

\begin{figure}[htbp]
\center
\resizebox{0.7\hsize}{!}{\includegraphics[angle=-90]{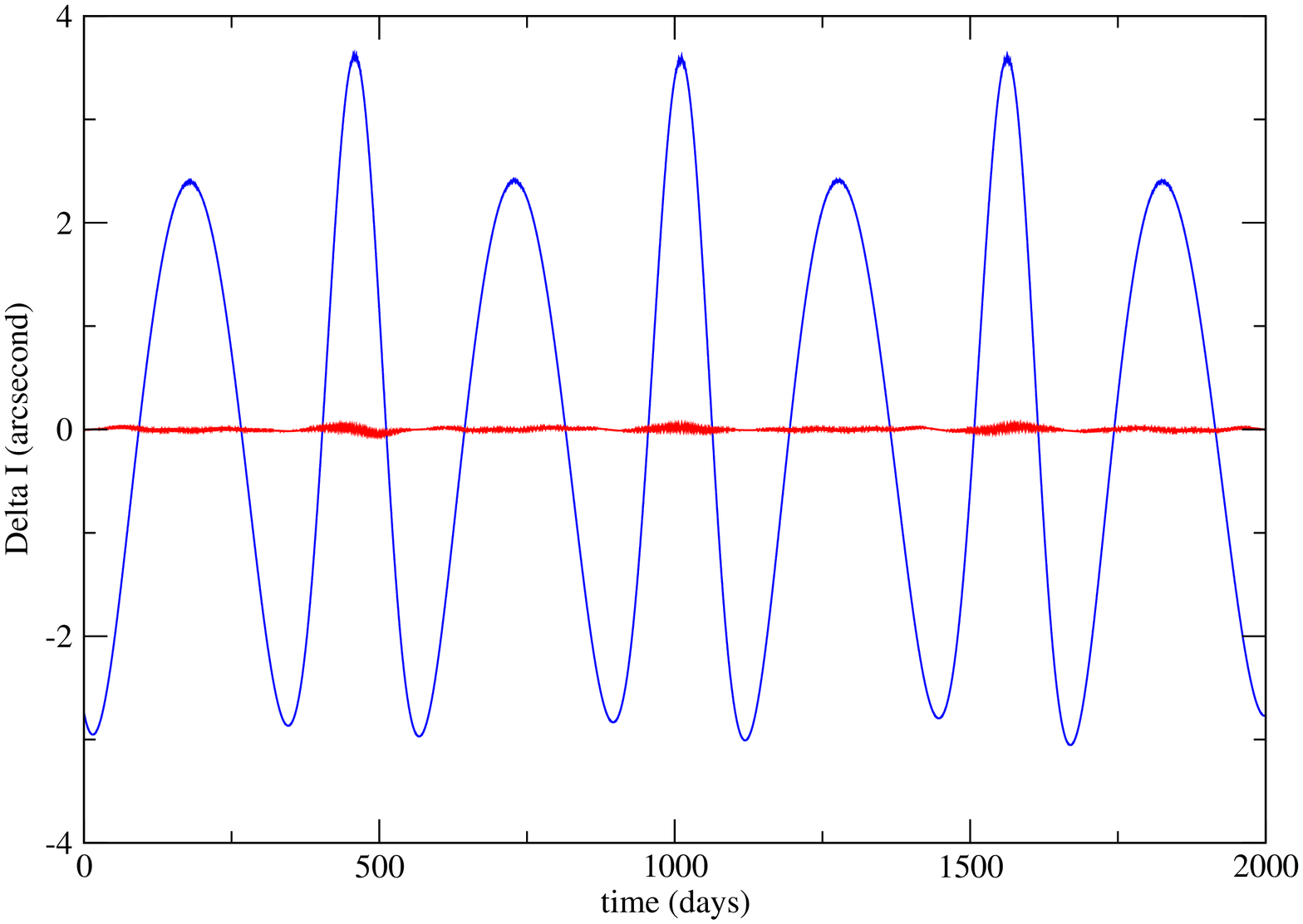}}
 \caption{The nutation of Phoebe in obliquity for a 2000 days times span.The central curve represents the residual after substracting the sinusoidal terms of the Table \ref{tab2}.}
\label{fig5}
\end{figure}

Using the ephemeris of Emelyanov (2007), the precession and the nutation in longitude and in obliquity are determined by numerical integration with an extrapolation-algorithm based on the explicit midpoint rule. We remind here that the precession and the nutation in longitude and in obliquity are defined as the variation in space of the angular momentum axis with respect to the reference axis due to Saturn's gravitational perturbation. They correspond respectively to the linear part of $h$ and the quasi-periodic part of $h$ and $I$. Before integration, the planeto-equatorial coordinates (X, Y, Z) taken from ephemerides are converted into rectangular coordinates with respect to the orbital plane, which at their turn are converted into the spherical coordinates used in (\ref{eq5}) and (\ref{eq6}). The values of the dynamical flattening and the triaxiality of Phoebe are taken in Aleshkina et al. (2010). Figs.\ref{fig3} and \ref{fig4} show respectively Phoebe precession-nutation in longitude and Phoebe nutation alone for a 9000 days time span (24.6 y). Fig.\ref{fig5} shows the nutation in obliquity for a 2000 days time span (5.47 y). Notice that 9000 days correspond to the limit of the ephemerides. The 2000 days time span of the nutation in obliquity is chosen in order to see clearly the leading oscillations. The precession and the leading oscillations of the nutation in longitude and in obliquity are determined thanks to a linear regression and a fast Fourier Transform (FFT). Our value obtained for the precession of Phoebe is :
\begin{equation}\label{precession}
\Psi =5580.65t
\end{equation}
where $t$ is counted in Julian centuries. Notice that this value for the precession in longitude is very close to the precession for the Earth ($\Phi=5000".3$cy). So the effect of the tidal torque of Saturn on the precession of Phoebe is roughly the same as the combined gravitational effect of the Moon and Sun on the precession of the Earth. The physical dissymmetry (large values of the dynamical flattening and of the triaxiality) of Phoebe which directly increase the amplitude of the precession are compensated by its slow revolution and fast rotation (see eq (\ref{eq9})) .

The nutation in longitude is given by :
\begin{eqnarray}\label{FFTlo}
\Delta \Psi &&= \sum \Phi_{i}\sin (\nu_{i}t) +\Phi'_{i}\cos (\nu_{i}t)+\Theta_{i} t \sin (\nu_{i}t)\nonumber\\&&
+\Theta'_{i} t \cos (\nu_{i}t)
\end{eqnarray}

where $\Phi_{i}, \Phi'_{i}, \Theta_{i}, \Theta'_{i}$ and $\nu_{i}$ are respectively the leading amplitudes and the frequencies of the sinusoidal terms and can be found in Table \ref{tab1}. Similarly, the nutation in obliquity is given by :
\begin{eqnarray}\label{FFTob}
\Delta I&& = \sum \Gamma_{i}\sin (\mu_{i}t) +\Gamma'_{i}\cos (\mu_{i}t)+\Pi_{i} t \sin (\mu_{i}t)\nonumber\\&&
+\Pi'_{i} t \cos (\mu_{i}t)
\end{eqnarray}
where $\Gamma_{i}, \Gamma'_{i}, \Pi_{i}, \Pi'_{i}$ and $\mu_{i}$ are given in Table \ref{tab2}. The corresponding arguments as a combination of Phoebe mean longitude and mean anomaly are determined empirically and given when they have been clearly identified. The residuals after substraction of (\ref{FFTlo}) and (\ref{FFTob}) to the results of the numerical integration are the flat curves in Figs.\ref{fig4} and \ref{fig5}.

\begin{table}[!h]
\caption{$\Delta \Psi=\Delta h $: nutation coefficients in longitude of Phoebe}
\label{tab1}
\begin{center}
\resizebox{.8\hsize}{!}{\begin{tabular}[htbp]{cccccc}
\hline \hline
 Argument &Period&$\sin$&$\cos$&$t\sin$&$t\cos$\\
 &$\frac{2\Pi}{\nu}$&$\Phi$&$\Phi'$&$\Theta$&$\Theta'$\\
  & days & arc second &arc second &arc second ($10^{-6}$)&arc second($10^{-6}$)\\
  \hline
$2L_{s}$ & 275.1353 & -3.411 & 5.8789&20.7222&-5.3705 \\

$2L_{S}-M$ &550.721&3.5260&4.29957&48.644&-44.6566 \\

$2L_{S}+M$ & 183.44&-2.4496& 0.2363&11.748&21.716\\

\ &5564.9&-0.2884&0.52829&-63.3917&31.786 \\

$2L_{S}+2M$  & 137.6& -0.482874&- 0.5341&-5.0143&10.129 \\

\  &497.34&-0.5300&0.14200& 18.758&8.77425\\

$2L_{S}+3M$&110.09&0.0348&-0.1723&-4.1343&-2.5833\\

$2L_{S}-5M$ & 177.17& 0.0413&-0.2188&-1.4336&7.4353\\

$2L_{S}-6M$&134.04&0.092656&-0.00384&-3.1435&0.243\\

\ &3220.0& 0.1203&0.1494&-1.53677&-30.596\\

\ &260.60&-0.1160&-0.0257&13.69&2.3207\\

\end{tabular}

}
\end{center}
\end{table}

\begin{table}[!h]
\caption{$\Delta \epsilon=\Delta I$ : nutation coefficients in obliquity of Phoebe}
\label{tab2}
\begin{center}
\resizebox{.8\hsize}{!}{\begin{tabular}[htbp]{cccccc}
\hline \hline
Argument &Period&$\sin$&$\cos$&$t\sin$&$t\cos$\\
 &$\frac{2\Pi}{\nu}$&$\Gamma$&$\Gamma'$&$\Pi$&$\Pi'$\\
  & days & arc second &arc second &arc second ($10^{-6}$)&arc second($10^{-6}$)\\
  \hline
$2L_{s}$ & 275.12& -2.266 & -1.4174&14786&1.2152 \\
$2L_{S}+M$ &183.45&-0.0937&-1.03465 &-9.01515&4.1209\\
$2L_{S}-M$ & 549.90&0.435& -0.1703&0.92524&3.7466\\
$2L_{S}+2M$  & 137.60&0.2326&-0.2053&-4.2893& -2.4026\\
$2L_{S}+3M$&110.09&0.0819&0.0164&-0.2074&-1.5682\\
$2L_{S}-5M$ & 177.15& 0.0910&0.01957&-3.3764&-0.7907\\
$2L_{S}-6M$&134.04&0.01659&0.0456&0.1052&-1.4259\\
\ &8166& -0.0089438&-0.1403&-5.4749&19.3857\\
\ &614&-0.025055&-0.02073&1.1722&-0.7703\\
$2L_{S}+4M$&91.75&0.0078745&0.01760&0.72270&-0.10350\\
&291.67&0.010783&-0.0228555&-1.17099&1.0911\\
&107.80&-0.007745&0.0123306&0.42834&-0.0620\\
&263.03&0.006117&-0.014634&-4.14544&3.24460

\end{tabular}

}
\end{center}
\end{table}

\begin{figure}[htbp]
\center
\resizebox{0.7\hsize}{!}{\includegraphics[angle=-90]{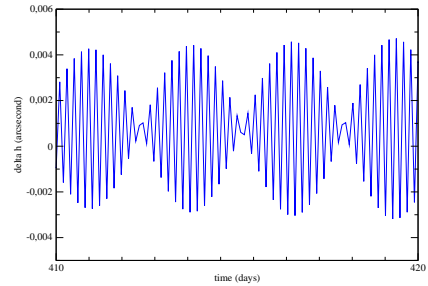}}
 \caption{Effect of the triaxiality on the nutation in longitude for a 10 days time span.}
\label{fig6}
\end{figure}

\begin{figure}[htbp]
\center
\resizebox{0.7\hsize}{!}{\includegraphics[angle=-90]{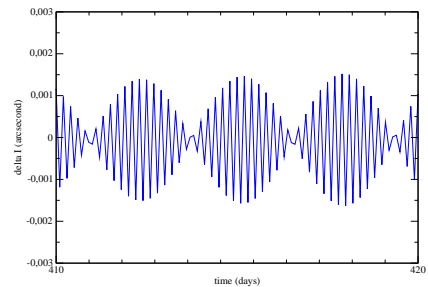}}
 \caption{Effect of the triaxiality on the nutation in obliquity for a 10 days time span.}
\label{fig7}
\end{figure}

The nutation in obliquity is dominated by two frequencies associated with the arguments $2L_{s}$ and $2L_{s}+M$ with respective periods 275.13 d and 183.45 d. In addition to these two frequencies, the nutation in longitude also presents another leading component with argument $2L_{s}-M$. The components which are not identified in Tables \ref{tab1} and \ref{tab2} as those with period 497 d and 260 d, already pointed out in Tables \ref{tab1new} and \ref{tab2new} (see Sect.\ref{3}). They probably come from the non Keplerian motion of Phoebe. As in the case for the precession, the nutations in longitude and in obliquity of Phoebe, which show respectively peak to peak variations of 26" or 8" are of the same order as the nutation of the Earth (respectively with a factor $\frac{2}{3}$ and $\frac{1}{2}$). As the value of the triaxiality is large (Aleshkina et al., 2010), it is interesting to evaluate its effect on the nutation of Phoebe. This effect is determined by the equations (\ref{eq5}) and (\ref{eq6}) taking only into account the component depending on $\frac{A-B}{4C}$ in equations (\ref{eq9}) and (\ref{eq10}). Figs \ref{fig6} and \ref{fig7} show respectively the nutation in longitude and in obliquity depending on the triaxiality for a 10 days times span beginning at 410 days. This short time span is chosen to see clearly the high frequency oscillation to the rotation of Phoebe with period roughly half the proper rotation of Phoebe (9 hrs) and the higher amplitude of the whole signal. Thus we remark that despite the large triaxiality of Phoebe, its effect on the nutation in longitude and in obliquity, at the level of a few $10^{-3}$ arcseconds amplitude, is very small . As explained already for the precession, the effect of the disymmetry of Phoebe, characterized by a rather large value of its triaxiality, is compensated by its very fast rotation, which strongly decreases the amplitude of the signal after integration.

The goal now is to determine an analytical model to describe the precession-nutation of Phoebe and to compare it with our numerical results. In the following the effect of the triaxiality on the nutation will be ignored and the $\frac{\mathtt{\textbf{G}}M'}{r^3}$ in (\ref{eq7}) and (\ref{eq8}) are replaced by $\frac{\mathtt{\textbf{G}}M'}{a^3}(\frac{a}{r})^3$.

\section{Construction of an analytical models for the precession-nutation of Phoebe}\label{6}

Many authors have elaborated an analytical model of the rotation of the Earth in order to understand the causes of the observed motion and to have the ability to predict it (Kinoshita, 1977, Bretagnon et al, 1997, Roosbeek and Dehant, 1997, Souchay et al., 1999). In this approach, it is interesting to build an analytical model based on the model of Kinoshita for the Earth to determine the precession and the nutation of Phoebe.

To solve analytically the (\ref{eq5}) and (\ref{eq6}), Kinoshita (1977) developed $\frac{1}{2}(\frac{a}{r})^3$ and $(\frac{a}{r})^3 \cos 2\lambda$ as a function of time, through the variables $M$ and $L_{s}$ taking the eccentricity as a small parameter :
\begin{eqnarray}\label{dl}
&&\frac{1}{2}(\frac{a}{r})^3 =f(M, e)+o(e^3)\nonumber\\&&(\frac{a}{r})^3 \cos 2\lambda =g(M,L_{s}, e)+o(e^3)
\end{eqnarray}
The detailed demonstration is presented in Kinoshita (1977). As the eccentricity of the Earth is small, truncating the development at the 3rd order is a sufficient approximation. On the opposite the eccentricity of Phoebe is higher (by one order) than the eccentricity of the Earth (0.18 instead 0.0167). Moreover, as we have seen before, Phoebe orbital elements are undergoing large variations. Then the relative error made by replacing $\frac{1}{2}(\frac{a}{r})^3$ and $(\frac{a}{r})^3 \cos 2\lambda$ by the developments such as (\ref{dl}) need to be quantified. To that purpose, $\frac{1}{2}(\frac{a}{r})^3$ and $(\frac{a}{r})^3 \cos 2\lambda$ are developed as for the Earth but with the real values of $e, M, L_{S}$ taken in the ephemerides of Emelyanov (2007), instead of a linear value (for $e, M$ ans $L_{s}$) for the Earth.

\subsection{Test for the developments of $\frac{1}{2}(\frac{a}{r})^3$ and $(\frac{a}{r})^3 \cos 2\lambda$.}

\begin{figure}[htbp]
\center
\resizebox{0.7\hsize}{!}{\includegraphics[angle=-90]{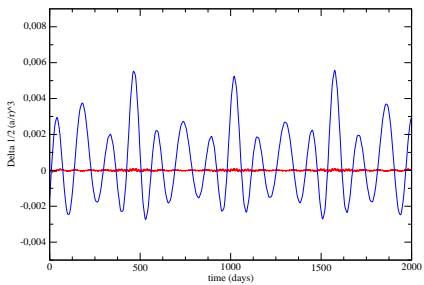}}
 \caption{$\Delta \frac{1}{2}(\frac{a}{r})^3$ between the numerical and the semi analytical results using the developments to 3rd (blue curve) and 6th order (red curve).}
\label{fig8}
\end{figure}

In this section, the developments of $\frac{1}{2}(\frac{a}{r})^3$ and $(\frac{a}{r})^3 \cos 2\lambda$ done by Kinoshita (1977) in the case of the Earth and by Cottereau and Souchay (2009) in the case of Venus are tested for Phoebe. We used the same methods as in Cottereau and Souchay (2009) to determine the development of $\frac{1}{2}(\frac{a}{r})^3$. As the eccentricity of Phoebe is large, the development at the 3rd order as made by Cottereau and Souchay (2009) for Venus is insufficient as it yields relative errors to the order of $4.10^{-3}$. This is due to the slow convergence of the development which must be carried out up to 6th order in our case, if we want to reach a relative $10^{-4}$ accuracy. Fig.\ref{fig8} shows the discrepancies between the numerical and the semi analytical results using this developments both to the 3rd and 6th order. This shows that high orders developments are needed when taking into account highly eccentric orbits as Phoebe's one.

\begin{figure}[htbp]
\center
\resizebox{0.7\hsize}{!}{\includegraphics[angle=-90]{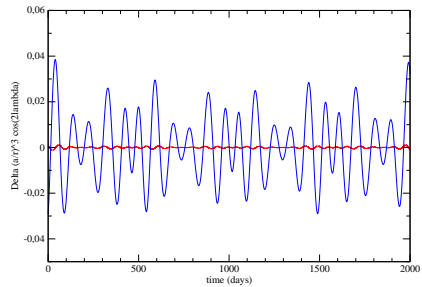}}
 \caption{$\Delta (\frac{a}{r})^3 \cos 2\lambda $ between the numerical and the semi analytical results using the developments to 3rd (blue curve) and 6th order (red curve).}
\label{fig9}
\end{figure}

For the developments of $(\frac{a}{r})^3 \cos 2\lambda$, the same conclusion is obtained. Fig.\ref{fig9} shows the discrepancies between the numerical and the semi analytical results using the developments of $\frac{a}{r}^3 \cos 2\lambda$, to 3rd and 6th order. The difference is descreased from 0".3 to 0".01 (peak to peak). The developments are given by :

\begin{eqnarray}\label{ar}
\frac{1}{2}(\frac{a}{r})^3&&=\frac{1}{2}+\frac{3}{4}e^2+\frac{15}{16}e^4+\frac{35}{32}e^6\nonumber\\&&
+(\frac{3}{2}e+\frac{27}{16}e^3+\frac{261}{128}e^5)\cos M \nonumber \\&&
+(\frac{9}{4}e^2 +\frac{7}{4}e^4+\frac{141}{64}e^6)\cos 2M\nonumber\\&&
+(\frac{53}{16}e^3+\frac{393}{256}e^5) \cos 3M\nonumber\\&&
+(\frac{77}{16}e^4+\frac{129}{160}e^6) \cos 4M
+\frac{1773}{256}e^5\cos 5M \nonumber\\&&+\frac{3167}{320}e^6 \cos 6M+O(e^7)
\end{eqnarray}

\begin{eqnarray}\label{arlam}
(\frac{a}{r})^3&& \cos 2 (\lambda-h) = (1-\frac{5}{2}e^2+\frac{13}{16}e^4+\frac{35}{288}e^6)\cos 2 L_{S}\nonumber\\&&
+\frac{4}{45}e^6 \cos (2 L_{S}-6M)+\frac{81}{1280}e^5\cos (2 L_{S}-5M)\nonumber\\&&
+(e^4 \frac{1}{24}+\frac{7}{240}e^6)\cos (2 L_{S}-4M)\nonumber\\&&
+(e^3 \frac{1}{48}+\frac{11}{768}e^5)\cos (2 L_{S}-3M)\nonumber\\&&
+(-\frac{1}{2}e^2+e^3 \frac{1}{16}-\frac{5}{384}e^5)\cos (2 L_{S}-M)\nonumber\\&&
+(\frac{7}{2}e^2+e^3 \frac{123}{16}+\frac{489}{128}e^5)\cos (2 L_{S}+M)\nonumber\\&&
+(\frac{17}{2}e^2+e^4 \frac{115}{6}+\frac{601}{48}e^6)\cos (2 L_{S}+2M)\nonumber\\&&
+(\frac{845}{48}e^3-e^5 \frac{32525}{768})\cos (2 L_{S}+3M)\nonumber\\&&
+(\frac{533}{16}e^4-e^6 \frac{13827}{160})\cos (2 L_{S}+4M)\nonumber\\&&
+\frac{228347}{3840}e^5\cos (2 L_{S}+5M)\nonumber\\&&
+\frac{73369}{720}e^6\cos (2 L_{S}+6M)+O(e^7)
\end{eqnarray}

\subsection{Analytical model of Phoebe}

\begin{figure}[htbp]
\center
\resizebox{0.6\hsize}{!}{\includegraphics[angle=-90]{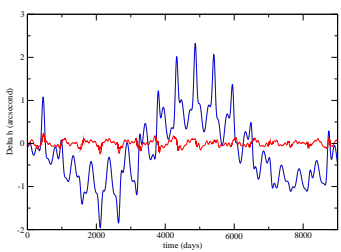}}
 \caption{Difference between the nutation in longitude of the analytical and numerical integration for 9000 days time span. The flat curve represents the residual after substraction (\ref{FFTlo}) to the nutation signal in longitude obtained by numerical integration.}
\label{fig12}
\end{figure}

\begin{figure}[htbp]
\center
\resizebox{0.6\hsize}{!}{\includegraphics[angle=-90]{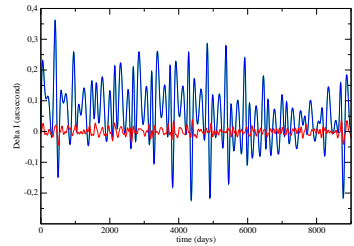}}
 \caption{Difference between the nutation in obliquity of the analytical and numerical integration for 9000 days time span. The flat curve represents the residual after substraction (\ref{FFTob}) to the nutation signal in obliquity obtained by numerical integration.}
\label{fig13}
\end{figure}

To express $\frac{1}{2}(\frac{a}{r})^3$ and $(\frac{a}{r})^3 \cos 2\lambda$ by purely analytical functions, Kinoshita replaced $e, M, L_{S}$ by their linear mean values. For Venus, the same method is used in Cottereau and Souchay (2009), who replaced $e, M, L_{S}$ by their linear mean values given by Simon et al (1994). Using the mean elements of the section (\ref{2}) and the developments given by (\ref{ar}) and (\ref{arlam}), an analytical model is obtained. As explained in Kinoshita and Souchay (1990) and Cottereau et al. (2010) for the Earth and for Venus, the difference between the nutation computed by analytical tables and the numerical integration is small (at the order of $10^{-5}$) and caused by the indirect planetary effects on the planet. For Phoebe, this difference is also caused by indirect perturbations but cannot be ignored as it will be presented in the following. Replacing the semi-major axis in (\ref{eq5}) and (\ref{eq6}) by $\bar{a}$ (Sect.\ref{3}), $\Delta h$ and $\Delta I$ depend directly on the scaling factor $K_{S}$ :
\begin{equation}
K_{S}=\frac{3n^2}{\omega}\frac{2C-A-B}{2C}=-11674".4857 \quad \mathtt{Julian.cy}
\end{equation} 
where $\frac{2C-A-B}{2C}$ is the dynamical flattening of Phoebe, $\omega$ the angular velocity of Phoebe and $n$ the mean motion of Phoebe given by :
\begin{equation}
n^2\bar{a}=GM'
\end{equation}
The numerical value is deduced directly from the values of the $A, B, C$ and $\omega$ taken respectively in Aleshkina et al. (2010) and Bauer et al (2004). Notice that this scaling factor is very close to the Earth one (11036".6 by Julian cy). As seen in the section (\ref{2}) the precession of Phoebe corresponds to the polynomial part of (\ref{eq5}) and can be directly determined by :
\begin{eqnarray}
\psi &&=-(K_{s}\cos I \int (\frac{1}{2}+\frac{3}{4}e^2+\frac{15}{16}e^4+\frac{35}{32}e^6) dt)
\nonumber\\&& =5541.18 t 
\end{eqnarray}
where $e$ is replaced by the linear part of (\ref{e}). The discrepancies between the precession as computed by analytical integration and that given by (\ref{precession}) are of the order of $1\%$ and show that the analytical model describes with a good accuracy the precession of Phoebe. Replacing respectively $e, M, L_{S}$ in the developments (\ref{ar}) and (\ref{arlam}) by the linear part of (\ref{e}),(\ref{mp}) and (\ref{ls2}), the nutation of Phoebe is determined by analytical integration. Figures \ref{fig12} and \ref{fig13} show the differences between the nutation in longitude and in obliquity of the angular momentum axis from a purely analytical formulation and numerical integration on 9000 days time span. These differences are compared to the residuals obtained in section \ref{5} after substraction (\ref{FFTlo}) and (\ref{FFTob}) determined by FFT to the numerical signal (central curves in the Figures \ref{fig12} and \ref{fig13}).

Figures \ref{fig12} and \ref{fig13} show that the semi-analytical functions characterizing the nutation of Phoebe respectively in longitude and in obliquity determined thanks to a fast Fourier Transform are more accurate than these obtained by a pure analytical integration (with a linear values of $e,M,L_{s}$). The pure analytical model restricts the expressions of the functions which describe the orbital elements. For the Earth (Kinoshita and Souchay, 1990) and for Venus (Cottereau et al. 2010), this model is very satisfying because their motion are close to a keplerian one and consequently the orbital parameters can be replaced by linear functions. As the motion of Phoebe is highly disturbed in comparison to the planets of the solar system, the curves of the temporal variations of the orbital elements (Emelyanov, 2007) cannot be fitted by simple functions as shown by the large variations around the mean elements in Sect.\ref{3}. Thus the analytical model does not describe the periodic components due to the indirect effect of the Sun or of the other satellites of Saturn. In contrast these effects are directly taken into account when using the functions given by (\ref{FFTlo}) and (\ref{FFTob}) because the FFT approach followed in Sect.\ref{5} is better suited to describe quasi-periodic motion. However the precession is less sensitive to these periodic variations, which are averaged during the calculation. This explains the similarity of the precession rate given by the two models. To conclude, the purely analytical model set by Kinoshita (1977) for the Earth gives a good first approximation of the nutation of Phoebe but a more sophisticated analytical model is needed if one wants a precision down to the $10^{-5}$ level. We precise here that this accuracy is internal to our model and is independant of the uncertainties in the observations. This shows that we are limited only by the uncertainties on the input data

\section{Conclusion}

The purpose of this paper was to calculate for the first time the combined motion of precession and nutation of 
the Saturn satellite Phoebe. 

One of the important steps in our study was to describe the orbital motion of Phoebe by fitting the curves of the temporal variations of the orbital 
elements $a, e, M$ and $L_{s}$ with polynomial functions. As shown in \ref{fig3new}, the periodic variations of the semi-major axis
around its mean value with a relative amplitude of the order of $10^{-3}$ are clearly due to the perturbing effects of other celestial bodies 
like the Sun. The orbital motion is far from being a keplerian one, as shown by the large polynomial expressions of $e, M$ and $L_{s}$, and by 
the large sinusoidal amplitudes characterizing the residuals after substraction of these polynomials.
  Applying the theoritical framework already used by Kinoshita (1977) for the Earth, the precession and the nutation motion of Phoebe 
are determined both analytically and from numerical integration of the equations of motion.
We found that the precession-nutation motion of Phoebe undergoing the gravitational perturbation of Saturn is quite similar to that the Earth 
undergoing the gravitational effect of both the Moon and the Sun.
Thus our value for the precession of Phoebe, that is to say 5580".65 cy, is very close to the corresponding value for the Earth (5081"/cy)
and the nutation in longitude and in obliquity of Phoebe with peak to peak variations of $26"$ and $8"$ are of the same order of amplitude as the nutation of the Earth (respectively 36" and 18" peak to peak).
Moreover Phoebe obliquity ($23^\circ.95$) is roughly the same as the Earth's one ($23^\circ.43$). Notice that 
the physical dissymmetry (large value of the dynamical flattening and of the triaxiality) and the large eccentricity of Phoebe 
which direclty increase the amplitude of the precession and the nutation is compensated by its slow revolution and fast rotation.

In Cottereau and Souchay (2009), we have demonstrated that the effects of the large triaxiality of Venus on its nutation are of the same order of magnitude than the effects of the dynamical flattening. By contrast, the effects of the large triaxiality of Phoebe on its nutation are compensated by its fast rotation and decrease the amplitude which become negligible compared to the dynamical flattening.

We also investigated the possibility to construct analytical tables of Phoebe nutation, as was done for the Earth.
As Phoebe has a large eccentricity, the analytical developments at the 3th order given by Kinoshita (1977) for the Earth are not valid.
We must carry out the developments up to 6th order to reach a relative $10^{-4}$ accuracy.

Thus although the amplitude of the nutation motion is close to the Earth one, we demonstrated that the analytical model used by Kinoshita 
for the Earth does not describe the nutation motion of a disturbed body like Phoebe with the same accuracy. This analytical model does not take into account the large perturbing effects of the celestial bodies on the orbit of the satellite.

To describe the nutation motion of Phoebe a FFT approach is better than a pure analytical integration done with linear expressions for $e, M$ and $L_S$ as shown Figs.\ref{fig12} and \ref{fig13}. The FFT analysis is better fitted to describe the periodic variations characterizing the nutation signals of Phoebe. These periodic variations are averaged during the calculation of the precession which decreases the discrepancies between the two models.

To conclude we have shown that the analytical model set by Kinoshita (1977) gives a good first approximation of the precession-nutation of Phoebe but further analytical developments are needed to reach the same accuracy than for the terrestrial planets.

We think that this work can be starting point for further studies such as the elaboration of another very precise analytical model
of the rotation of Phoebe by taking into account effects ignored in this paper, as the direct effects of the Sun, of Titan and of Saturn dynamical flattening. Such a model is required to develop the long term ephemerides of Phoebe's rotation, which should require long term orbital ephemerides, not still available.

\section{Appendix I}
\begin{table}[!h]
\caption{Numericals values used in this paper}
\label{table1}
\begin{center}
\resizebox{0.8\hsize}{!}{\begin{tabular}[h]{lc}
\hline \hline\\
  & Phoebe \\
\hline\\
Eccentricity (J2000.0) &0.1648 \\
\hline\\
Inclination (J2000.0) & $151^{\circ}.64$\\
\hline\\
Ascending node (J2000.0) &$54^{\circ}.317$ \\
\hline\\
Obliquity & $23^{\circ}.95$ \\
\hline\\
Period of rotation &0.386396 d \\
\hline\\
 Triaxiality : $\frac{A-B}{4C}$ &-0.011125 \\\\
\hline\\
 Dyn. flattening : $\frac{2C-A-B}{2C}$ &0.06465 \\\\
\hline\\
geo-equatorial coordinates of the pole of Phoebe &$\alpha_{p}^{0} =356^{\circ}.90$ and $\delta_{p}^{0} =77^{\circ}.80$\\\\
\hline\\
geo-equatorial coordinates of the pole of Saturn &$\alpha_{S0} =40^{\circ}.589$ and $\delta_{S0}=83^{\circ}.537$\\\\
\end{tabular}
}
\end{center}
\end{table}


\begin{thebibliography}{}

\bibitem{} Aleshkina, E.~Y., 
Devyatkin, A.~V., \& Gorshanov, D.~L.\ 2010, IAU Symposium, 263, 141 

\bibitem{} Andoyer H.\ 1923, Paris, 
Gauthier-Villars et cie, 1923-26.

\bibitem{} Bauer, J.~M., Buratti, 
B.~J., Simonelli, D.~P., \& Owen, W.~M., Jr.\ 2004, \apjl, 610, L57 

\bibitem{} Bretagnon P., Rocher P., \& Simon J.~L.\ 1997, \aap, 319, 305 ,

\bibitem{}  Capitaine N., 
Souchay J., \& Guinot B.\ 1986, Celestial Mechanics, 39, 283

\bibitem{} Cottereau, L., Souchay, J., \& Aljbaae, S.\ 2010, \aap, 515, A9 

\bibitem{} Cottereau, L., \& Souchay, J.\ 2009, \aap, 507, 1635

\bibitem{} Emelyanov, N.~V.\ 2007, \aap, 473, 343 

\bibitem{} Jacobson, R.~A.\ 1998, \aaps, 128, 7 

\bibitem{} Kinoshita H.\ 1977, 
Celestial Mechanics, 15, 277 

\bibitem{} Kinoshita, H., \& Souchay, J.\ 1990, Celestial Mechanics and Dynamical Astronomy, 48, 187 

\bibitem{} Meyer, J., \& Wisdom, J.\ 2008, Icarus, 193, 213 

\bibitem[]{} Rambaux, N., Lemaitre, A., \& D'Hoedt, S.\ 2007, \aap, 470, 741 

\bibitem{} Roosbeek, F., \& Dehant, V.\ 1997, IAU Joint Discussion, 3, 

\bibitem{} Ross, F.~E.\ 1905, Annals of 
Harvard College Observatory, 53, 101 

\bibitem{} Seidelmann, P.~K., 
et al.\ 2007, Celestial Mechanics and Dynamical Astronomy, 98, 155 

\bibitem{} Simon J.~L., Bretagnon P., Chapront J., Chapront-Touze M., Francou G., \& Laskar J.\ 1994, \aap, 282, 663

\bibitem{} Sinclair, A.~T.\ 1977, 
\mnras, 180, 447 


\bibitem{} Souchay J., Loysel B., Kinoshita H., \& Folgueira M.\ 1999, \aaps, 135, 111 

\bibitem{} Wisdom, J., Peale, 
S.~J., \& Mignard, F.\ 1984, Icarus, 58, 137 

\bibitem{} Wisdom, J.\ 1987, \aj, 94, 1350

\bibitem{} Woolard, E.~W.\ 1953, 
Astronomical papers prepared for the use of the American ephemeris and 
nautical almanac, v.~15, pt.~1, Washington, 1953 

\bibitem{} Yoder, C.~F., \& Standish, E.~M.\ 1997, \jgr, 102, 4065

\end{thebibliography}
\end{document}